\documentclass[twocolumn, times, tighten]{aastex62}
\pdfoutput=1 

\usepackage{natbib}
\usepackage{amssymb}
\usepackage{amsmath}
\usepackage{amstext}
\usepackage{bm}
\usepackage{color}
\usepackage{graphicx}
\usepackage{graphicx}
\usepackage{color}
\defcitealias{fitz19}{Paper VII}

\newcommand{\rv}{$R(V)$}

\newcommand{ \invmic}{$\mu^{-1}$}
\newcommand{\ebv}{$E(B-V)$}

\received{January 1, 2018}
\revised{January 7, 2018}
\accepted{\today}
\submitjournal{ApJ}

\shorttitle{Optical extinction}

\shortauthors{Massa, Fitzpatrick \& Gordon}

\begin{document}
\title{An Analysis of the Shapes of Interstellar Extinction Curves. VIII. The 
Optical Extinction Structure
\footnote{Based on observations made with the NASA/ESA Hubble Space Telescope, 
obtained at the Space Telescope Science Institute, which is operated by the 
Association of Universities for Research in Astronomy, Inc., under NASA 
contract NAS 5-26555. These observations are associated with program \# 
13760.}}

\correspondingauthor{Derck Massa}
\email{dmassa@spacescience.org}

\author[0000-0002-9139-2964]{Derck Massa}
\affil{Space Science Institute, 4750 Walnut St., Suite 205, Boulder, CO 
80301, USA}

\author[0000-0002-2371-5477]{E.\ L.\ Fitzpatrick}
\affiliation{Department of Astronomy \& Astrophysics, Villanova University, 
800 Lancaster Avenue, Villanova, PA 19085, USA}

\author[0000-0001-5340-6774]{Karl D.\ Gordon}
\affiliation{Space Telescope Science Institute, 3700 San Martin
  Drive, Baltimore, MD, 21218, USA}
\affiliation{Sterrenkundig Observatorium, Universiteit Gent,
  Gent, Belgium}

\begin{abstract}
New {\it HST}/STIS optical spectra were obtained for a sample of early type 
stars with existing {\it IUE}\/ UV spectra.  These data were used to 
construct optical extinction curves whose general properties are discussed 
elsewhere.  In this paper, we identify extinction features in the curves 
that are wider than diffuse interstellar bands (DIBs) but narrower than the 
well known broad band variability.  This intermediate scale structure, or 
ISS, contains distinct features whose peaks can contribute a few percent to 
20\% of the total extinction.  Most of the ISS variation can be captured by 
three principal components.  We model the ISS with three Drude profiles and 
show that their strengths and widths vary from one sight line to another, 
but their central positions are stable, near 4370, 4870 and 6300~\AA.  The 
Very Broad Structure, VBS, in optical curves appears to be a minimum between 
the 4870 and 6300~\AA\ absorption peaks.  We find relations among the fit 
parameters and provide a physical interpretation of them in terms of a 
simplistic grain model.  Finally, we note that the strengths of the 4370 and 
4870~\AA\ features are correlated to the strength of the 2175~\AA\ UV bump, 
but that the 6300~\AA\ feature is not, and that none of the ISS features are 
related to $R(V)$.  However, we verify that the broad band curvature of the 
continuous optical extinction is strongly related to $R(V)$. 
\end{abstract}


\section{Introduction\label{sec:intro}} 
It has been known for some time that the broad band structure of optical and 
near infrared (NIR) extinction curves varies with location in the Galaxy 
\citep[see][for a review]{2016ApJ...821...78S}.  Likewise, spatially-variable 
narrow band extinction features, called diffuse interstellar bands (DIBs), 
have also been studied extensively \citep[e.g.][]{1995ARA&A..33...19H}.  In 
contrast, weak structure in optical and NIR extinction curves over 
intermediate wavelength intervals (i.e., several hundred to $\sim$1000~\AA) 
is known to exist, but has received far less attention.  This structure was 
first reported by \cite{1966ApJ...144..305W} and termed ``very broad 
structure'', or VBS.  The VBS is typically identified as a broad depression 
(i.e., reduced extinction) in extinction curves over the region $1.5 \lesssim 
\lambda^{-1} \lesssim 2.0$ \invmic\ \citep{1976MNRAS.177..625W, 
1977AJ.....82..337S, 1980PASP...92..411W}.  However, \cite{1971ApJ...166...65Y} 
and \cite{1973IAUS...52...83H} recognized that it could also be a minimum 
associated with larger, more complex structure.  In this paper, we 
collectively refer to these features, including the VBS, as intermediate scale 
structure, or ISS. Interestingly, there is little evidence for DIBs or ISS in 
the UV or far UV \citep[e.g.][]{2003ApJ...592..947C, 2009ApJ...705.1320G}.

Recently, \citet[][hereafter Paper VII]{2019ApJ...886..108F} have constructed 
\defcitealias{2019ApJ...886..108F}{Paper VII}
a set of high signal-to-noise spectrophotometric extinction curves which are 
ideal for studying the nature of the relatively weak ISS.  In this paper, 
we utilize the \citetalias{2019ApJ...886..108F}  data to examine the nature and 
variability of the ISS in detail.  We do this using two distinct 
approaches.  The first is purely empirical and does not rely on any 
assumptions about the form of the structure.  The second approach uses a 
parameterization of the curves in order to quantify the features 
identified in the empirical analysis. 

Section~\ref{sec:sample} describes our sample of stars and the data used in 
the analysis.  Section~\ref{sec:measure} presents our empirical measurements 
of the ISS.  Section~\ref{sec:modeling} utilizes a simple, {\it ad hoc} 
model for the structure in order to derive physically meaningful 
measurements.  Finally, section~\ref{sec:summary} summarizes our findings. 

\section{The Sample}\label{sec:sample}
We begin with the sample of 72 early-type stars described in 
\citetalias{2019ApJ...886..108F}. These stars all have HST/STIS G430L and 
G750L spectra, covering the wavelength range $2900 \leq \lambda \leq 
10270$~\AA, with a resolution ($\lambda/\Delta\lambda$) ranging from 530 to 
1040.  Additionally, all have been observed previously by the {\it 
International Ultraviolet Explorer} {\it IUE} satellite, providing 
ultraviolet (UV) spectrophotometry.  The stars range in spectral type from B9 
to O6 and luminosity class from V to III. They have color excesses spanning 
the range $0.12 \leq E(B-V) \leq 1.11$ mag.  \citetalias{2019ApJ...886..108F} 
derived extinction curves for each star using a procedure that matches stellar 
lines to model atmospheres.  The method is similar to the one employed by 
\cite{2007ApJ...663..320F} except that it concentrates on the optical spectrum 
\citepalias[see][for details]{2019ApJ...886..108F}.  We excluded 
GSC03712-01870 from this original sample because it lacked an {\it IUE}\/ 
long wavelength spectrum and, thus, information on the strength of its 
2175~\AA\/ extinction bump.  This left a sample of 71 stars.   

For our analysis, we rebinned all of the optical curves for these stars 
onto a uniform wavelength scale, covering $3010 \leq \lambda \leq 8000$~\AA\ 
and sampled at 5~\AA\ intervals.  Wavelengths longer than 8000~\AA\ were 
ignored since the available TLUSTY model atmospheres \citep{2003ApJS..146..417L} 
used to produce many of the curves do not include the upper Paschen lines.  
The final data set thus consists of 71 normalized extinction curves.  These 
curves are expressed as color excesses relative to the monochromatic flux at 
5500~\AA, $E(\lambda -55)$ and normalized by the excess between 4400 and 
5500~\AA, $E(44-55)$, i.e., 
\begin{equation}
k(\lambda -55) \equiv E(\lambda -55)/E(44-55)
\end{equation}
sampled at $M_\lambda = 999$ evenly spaced wavelength points.  These 
monochromatic excesses and curves are nearly identical to those expressed 
in the usual photometric $B$ and $V$ bands.  The major difference is that 
the monochromatic expressions are immune to the magnitude of the reddening 
\citepalias[see][for further details]{2019ApJ...886..108F}.  Nevertheless, 
we will often use  ebv\ or $R(V)$ as general measures of the color excess 
and the ratio of total to selective extinction. 

\begin{figure*}
\begin{center}
\includegraphics[width=1.0\linewidth, angle=180]{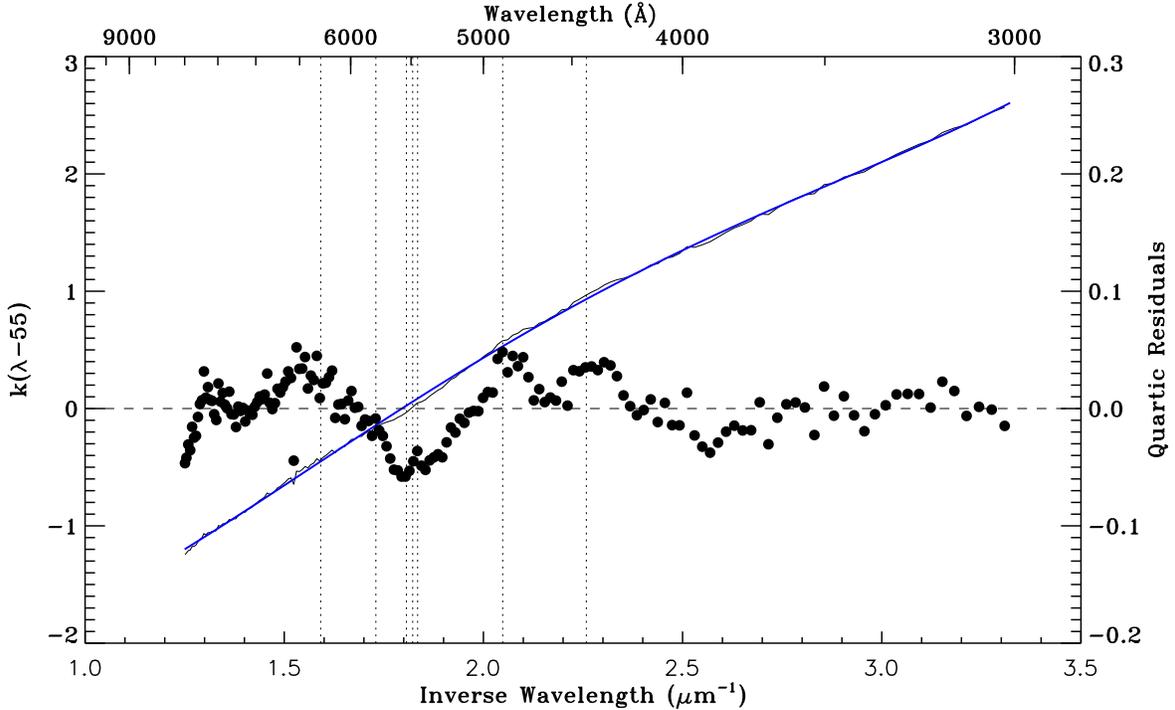}
\end{center}
\vspace{-0.75in}
\caption{Mean of 71 MW curves determined by \citetalias{2019ApJ...886..108F} 
and described in detail in \S \ref{sec:sample}, along with a quartic fit 
(blue curve) to the mean and $10\times$ the mean of the residuals.  The 
residuals for individual curves in the sample can be nearly twice as large 
or almost nonexistent. The vertical dotted lines give the positions of the 
strong DIBs listed in Table~\ref{tab:ism} (given in \S~\ref{sec:measure}).
\label{fig:mean_quartic}}
\end{figure*}

Figure~\ref{fig:mean_quartic} shows the mean of the 71 rebinned, normalized 
optical extinction curves, along with a quartic fit to the mean (smooth blue 
curve).  To demonstrate the magnitude of the features to be studied in this 
paper, we also show the difference between the mean curve and the quartic 
fit, magnified by a factor of ten. These residuals reveal the ISS.  

\section{Empirical Measurements of the ISS}\label{sec:measure}
In this section, we fit the curves with a low order polynomial.  We then 
examine the residuals of the polynomial fits, i.e., the ISS.  We analyze the 
variability of the residuals in two ways.  First, we quantify its strength 
and demonstrate that it is related to the amplitude of the 2175~\AA\ bump 
\citep[a result anticipated by][]{1980PASP...92..411W}. Second, we analyze 
variations in the shape of the ISS and show that most of it can be captured 
by only three parameters. 

To begin, we interpolate the curves over the locations of interstellar 
lines and those DIBs from the list given by \cite{1995ARA&A..33...19H} that 
we could identify in our spectra.  This allows our analysis to concentrate 
only on ISS features.  The interstellar features removed are listed in 
Table~\ref{tab:ism}. The curves were then smoothed by a Gaussian with a 
FWHM of 100~\AA.  Smoothing accomplishes two things: 1) it suppresses 
features with characteristic widths less than 100~\AA\ (eliminating any 
residual influence of DIBs) and, 2) it reduces the ``offset error'' 
described below.  

We can also use the smoothed and unsmoothed data to derive a measure of 
the mean error affecting the data.  First, we subtract each 999 point 
spectrum with the ISM features removed from the version of itself 
convolved with the 100~\AA\ FW Gaussian.  This difference should capture 
most of the high frequency, random noise in the data.  Next, the RMS of 
these differences at each wavelength for all of the observations is 
used to provide a representation of the wavelength dependence of the 
errors for the entire sample.  Finally, the RMS of this wavelength 
array is used to derive a ``representative'', mean error affecting the 
data.  The value that results from this process is $\sigma_{obs} = 0.0373$. 


\begin{table}[h]
\begin{center}
\caption{Interstellar Features}
\begin{tabular}{lcc}
\label{tab:ism}
Type  & $\lambda$ (\AA)  & $\lambda^{-1}$ (\invmic)\\ \hline
Ca II & 3933.66 & 2.54 \\
Ca II & 3968.47 & 2.52 \\
Na I  & 5889.95 & 1.70 \\
Na I  & 5895.92 & 1.70 \\
K I   & 7664.91 & 1.30 \\
K I   & 7698.97 & 1.30 \\ \hline
DIB   & 4428.00 & 2.26 \\
DIB   & 4882.00 & 2.05 \\
DIB   & 5450.30 & 1.83 \\
DIB   & 5487.50 & 1.82 \\
DIB   & 5535.00 & 1.81 \\
DIB   & 5780.45 & 1.73 \\
DIB   & 6283.86 & 1.59 \\ \hline
\end{tabular}
\end{center}
\end{table}

\subsection{Magnitude of the ISS}
We assume that the curves can be represented by a set of $N_f$ feature 
profiles, $\phi(\lambda -55)_i$, whose shapes are unspecified, and a 
background continuum that can be characterized by a polynomial of order 
$N_p$.  In this case, an extinction curve can be expressed as 
\begin{equation}
k(\lambda -55) = \sum_{i=1}^{N_f} a_i^0 \phi(\lambda -55)_i + 
                 \sum_{j=0}^{N_p} b_j^0 \lambda^j 
\label{eq:flux}
\end{equation}

Without model profiles for guidance, weak, broad structures, such as the 
ISS, are difficult to quantify.  Unlike sharp features, such as DIBs, 
choosing continuum points is problematic.  It is difficult to decide where 
the ``continuum'' begins and ends and whether a perceived continuum point 
is actually affected by the overlap of nearby features.  In the end, any 
such procedure becomes highly subjective.  Because of these difficulties, we 
begin our analysis with a different approach for measuring the ISS.  Instead 
of attempting to fit the features with a set of profiles or to determine the 
underlying continuum, we examine the magnitudes of the residuals of the 
curves after they are fit by low order polynomials.  Specifically, the 
residuals, $r(\lambda-55)$, are given by 
\begin{equation}
r(\lambda -55) = k(\lambda-55) - \sum_{j=0}^{N_p} b_j \lambda^j
\label{eq:resid}
\end{equation}
where the $b_j$ differ from the $b^0_j$ because they are fits to the entire 
curve, including the ISS.  Consequently, they are affected by the ISS at 
some level.  Experimentation showed that a fourth order polynomial (a 
quartic) adequately captures the overall shape of the curves and that higher 
order polynomials do not improve the fits.  

After the smoothing and the removal of a quartic continuum, 
the features in the residuals have full widths between about 100 and 1200~\AA, 
where the upper limit is the total wavelength interval divided by 4 (for 
the 4 zeros of a 4-th order polynomial).  Features larger than 1200~\AA\ 
are captured by the polynomial coefficients, while features less than 
100~\AA\ are lost, but should stand out in the raw spectra (the ISM lines 
and DIBs).  As expected from \citetalias{2019ApJ...886..108F}, the 
coefficients of the linear and quadratic terms are correlated with $R(V) 
\equiv A(V)/E(B-V)$, and we return to this correlation in \S~\ref{sec:relations}.

\begin{figure}
\begin{center}
\includegraphics[width=1.0\linewidth]{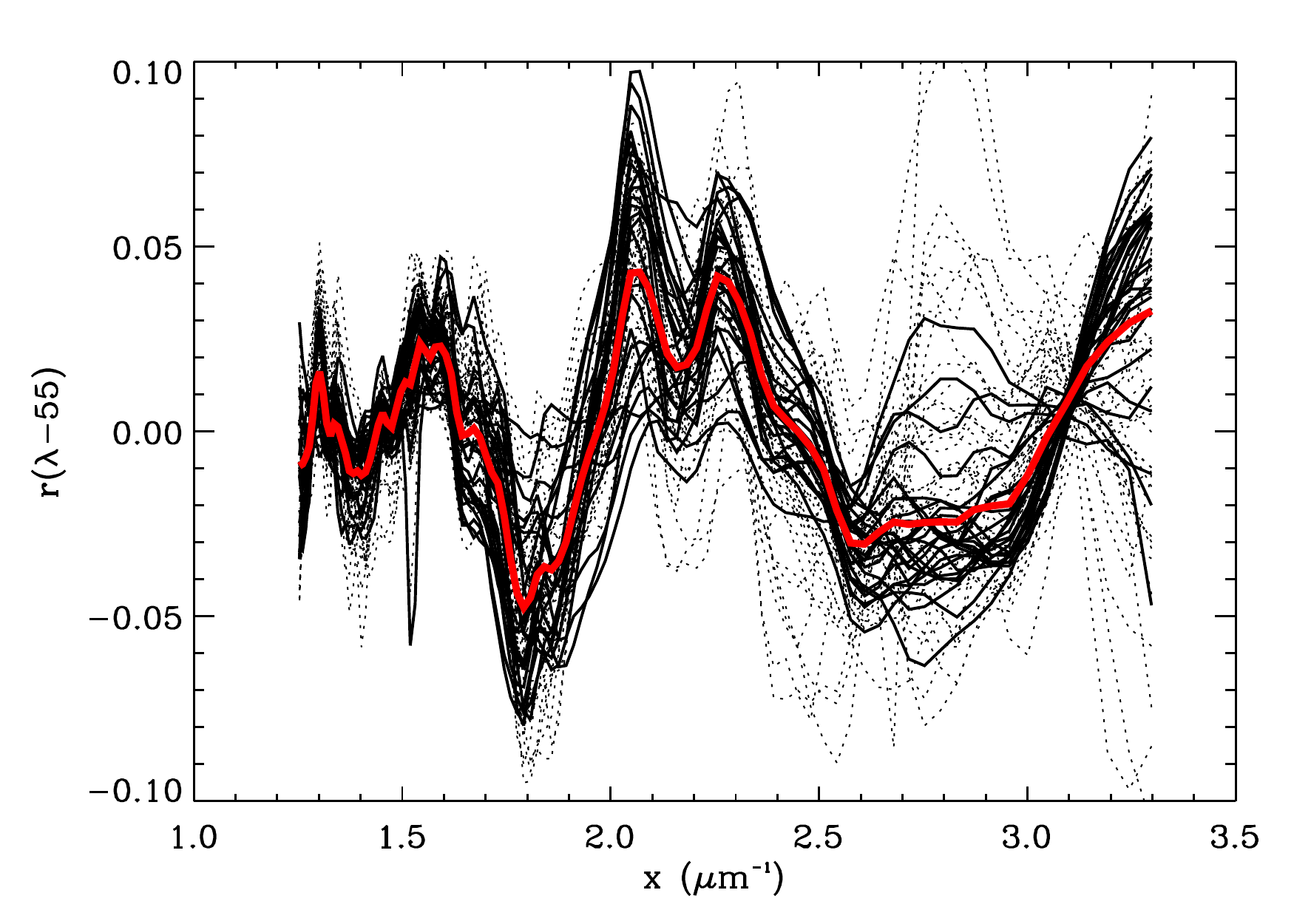}
\end{center}
\caption{Individual $r(\lambda-55)$ curves (black) and their mean (red) 
plotted against inverse wavelength.  Dotted and solid lines are used to 
denote curves derived from stars with $E(44-55)$ less than or greater than 
0.5 mag, respectively.  
\label{fig:kperps}}
\end{figure}

Figure~\ref{fig:kperps} shows the individual residuals and 
their mean.  The dotted curves are for stars with $E(44-55) < 0.5$ mag, 
where the effect of mismatches between the observations and the best 
fitting model atmospheres used to create the curves are exaggerated 
\citep[see][for a discussion of mismatch errors]{1983ApJ...266..662M}.  
A few things are immediately apparent.  First, the general structure is 
repeatable and present at some level in all of the well defined curves.  
Second, the broad depression in the extinction between $1.5 \lesssim x 
\lesssim 2.0$ \invmic (where $x \equiv \lambda^{-1}$), which is normally 
associated with the VBS, is clearly present.  Third, the VBS depression is 
actually a minimum surrounded by at least three distinct peaks, near $x = 
1.6$, 2.05, and 2.25 \invmic ($\lambda = $4400, 4800 and 6600~\AA).  This 
is the same general structure first reported by \cite{1973IAUS...52...83H}.  
Finally, we note that the three distinct peaks in the ISS occur in the 
vicinity of the three strong DIBs at 4428, 4882 and 6384~\AA\ (see 
Table~\ref{tab:ism}). 

The repeatability of the features validates the approach used in 
\citetalias{2019ApJ...886..108F} to construct the curves, since the same 
features appear in curves derived from stars with very different physical 
parameters and a wide range in color excesses -- indicating that these 
features are not a result of spectral mismatch between observed spectra 
and model flux calculations. The large scatter between $x = 2.7$ and 2.9 
\invmic\ results from small mismatches between the models and the 
observations in the vicinity of the Balmer Jump, which are amplified in 
curves produced from stars with smaller \ebv\ values.  It appears that 
the models or our fitting procedures may need to be refined in this region.  

\begin{figure}
\begin{center}
\includegraphics[width=0.9\linewidth]{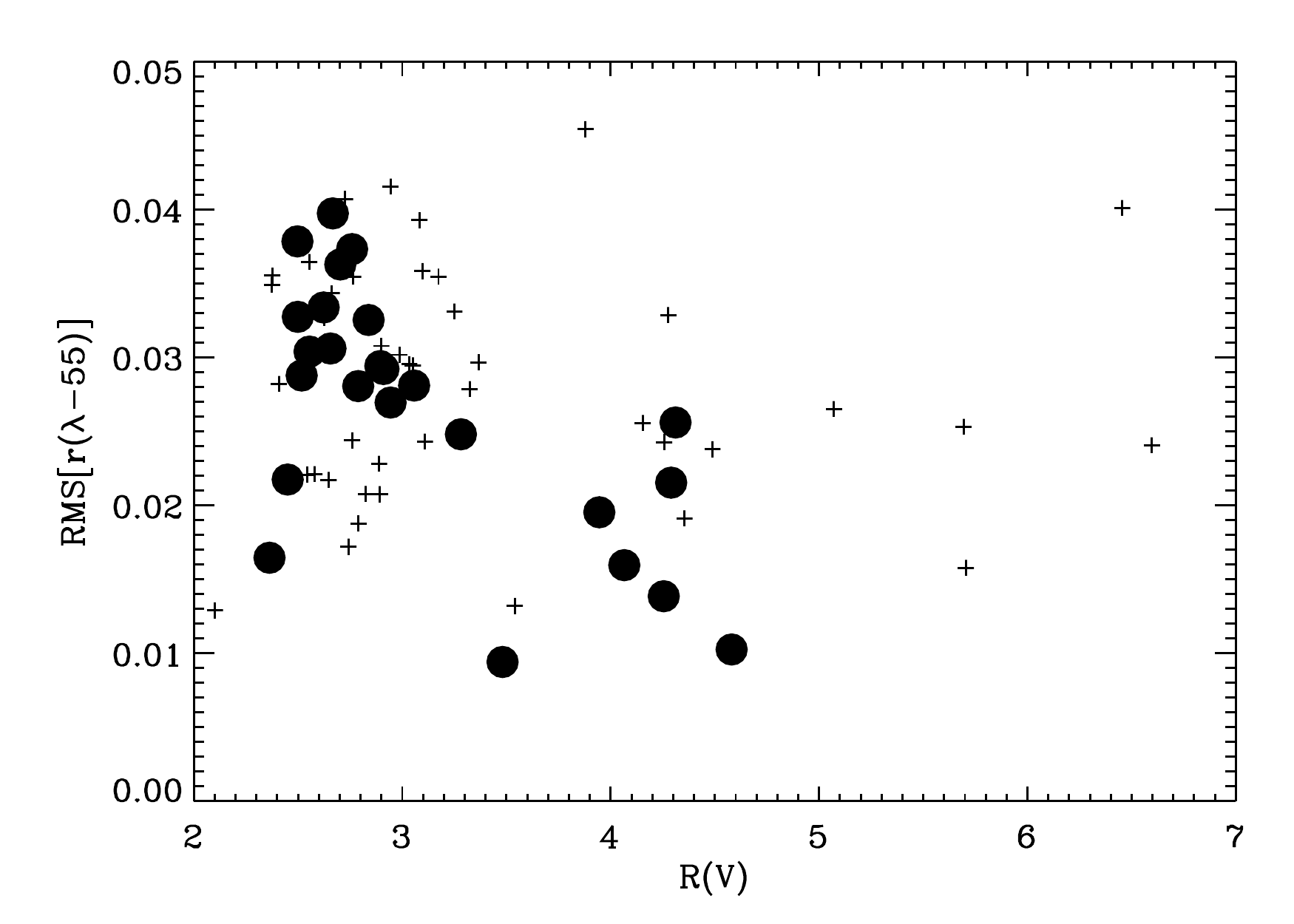}\vfill
\includegraphics[width=0.9\linewidth]{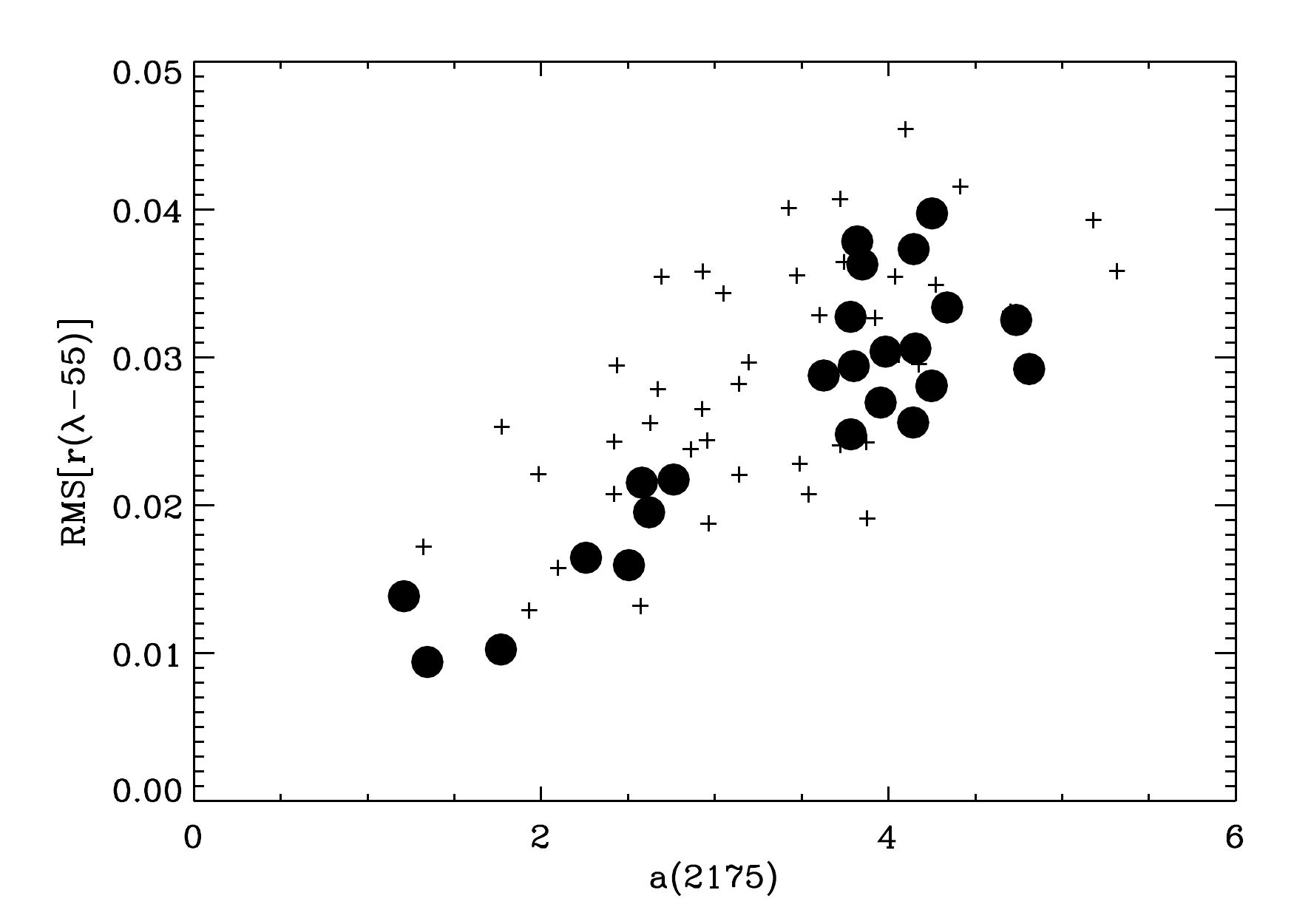}
\end{center}
\caption{$RMS[r(\lambda-55)]$ plotted against $R(V)$ (top) and bump 
amplitude, $a(2175)$\ (bottom).  Large, filled symbols are for values 
derived from curves with $E(44-55) \geq 0.5$ mag, which should be most 
accurate.  
\label{fig:bump}}
\end{figure}

We now require an objective means to measure the strength of the ISS.  For 
lack of a better measure, we adopt the root mean square (RMS) of the 
residuals to measure its total strength.  This is given by 
\begin{equation}
RMS[r(\lambda-55)] \equiv \sqrt{\frac{1}{M_\lambda}\sum_{k=1}^{M_\lambda} 
r(\lambda_k-55)^2} 
\label{eq:meas}
\end{equation}
where $M_\lambda = 999$.  Note that $RMS[r(\lambda-55)]$ is also influenced 
by random, point to point measurement errors which create an additive term 
under the radical.  This term creates an ``offset error'' (referred to above), 
in the sense that the measure can never be exactly zero, due to the random 
contribution.  Further, in very noisy data the random errors can dominate 
the measurements.  This was the motivation for aggressively smoothing the 
data with a 100~\AA\ Gaussian, which should minimize the random contribution.
Without the smoothing, there would be an additional term under the 
radical in equation~(\ref{eq:meas}) of order $\sigma_{obs}$.  This term 
would introduce considerable scatter in Figure~\ref{fig:bump} and prohibit 
curves with no features whatsoever from approaching zero.


We next examine correlations between $RMS[r(\lambda-55)]$ and $R(V)$ 
and properties of the 2175~\AA\ bump.  \cite{1986ApJ...307..286F} 
parameterize the 2175~\AA\ bump with a constant, $c$, times a Drude profile 
of the form 
\begin{equation}
D(x, x_i, \gamma_i) = \frac{x^2}{(x^2 - x_i^2)^2 +x^2\gamma_i^2}
\label{eq:drude}.
\end{equation}
The maximum value, or amplitude, of the term is $a(2175) = c/\gamma^2$ and 
its area is $\pi c/(2 \gamma)$ \citep[see][]{1986ApJ...307..286F}.  

Figure~\ref{fig:bump} shows the relations between $RMS[r(\lambda-55)]$ and 
\rv\ (top) and $a(2175)$ (bottom).  Values of $a(2175)$, and $R(V)$ are 
from \citetalias{2019ApJ...886..108F}.  It is clear that $RMS[r(\lambda-55)]$ 
is poorly correlated with $R(V)$.  In contrast, our measure of the ISS 
structure is strongly correlated with $a(2175)$.  Considering the crudeness 
of our current measure, this seems rather remarkable.  

\begin{figure}
\begin{center}
\includegraphics[width=1.0\linewidth]{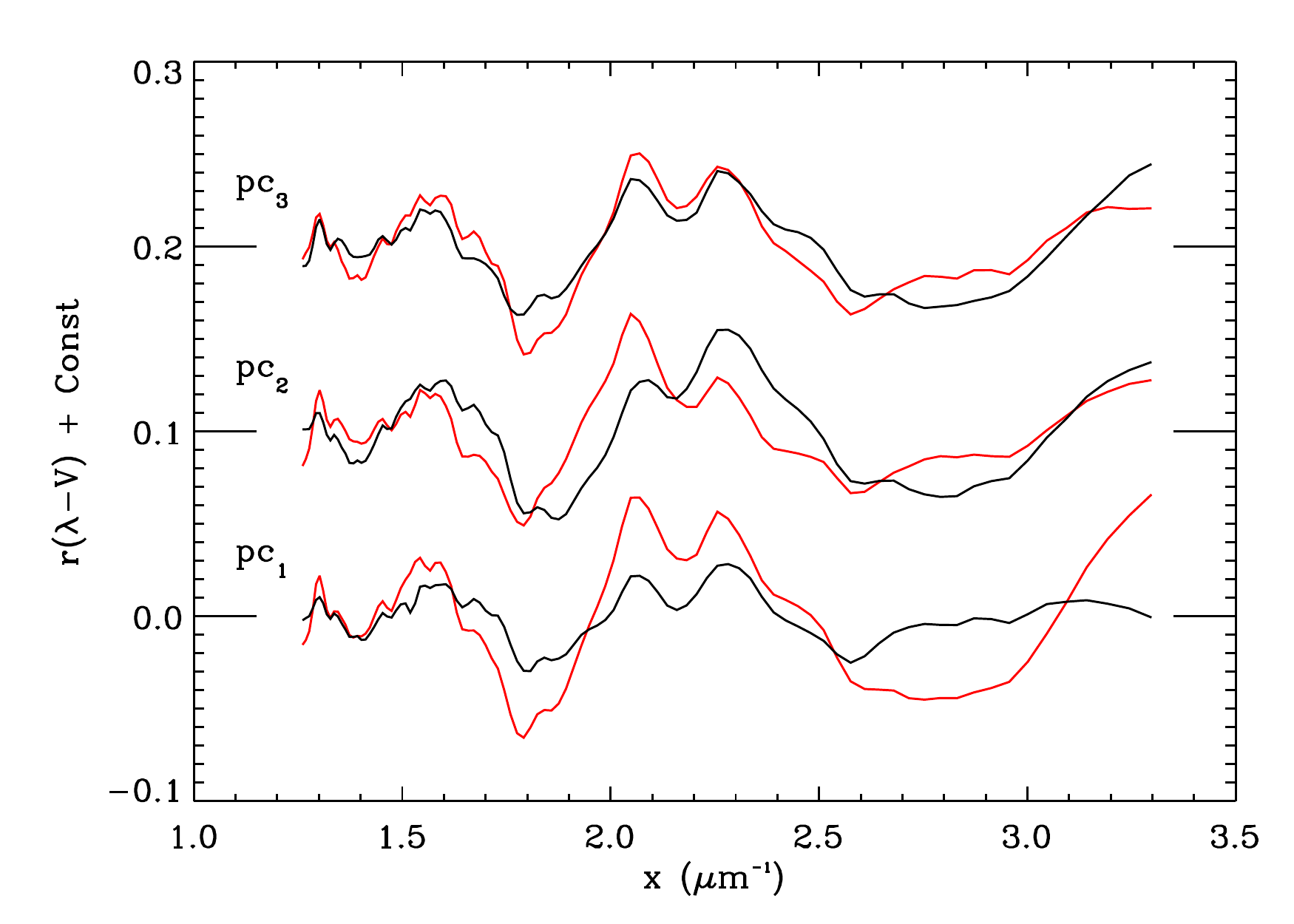}
\end{center}
\vspace{-0.2in}
\caption{The effects of the three largest principal components of 
$r(\lambda-55)$ on the sample mean plotted against inverse wavelength.  Each 
pair of curves shows the effect of adding (black) and subtracting (red) the 
PC to the mean.  The zero point of each set of curves is shown by the large 
tick marks and the curves are arranged with the largest PC at the bottom.  
The variations caused by the PCs are scaled by the magnitude of their 
variances.
\label{fig:pcs}}
\end{figure}
\begin{figure*}
\begin{center}
\includegraphics[width=1.0\linewidth]{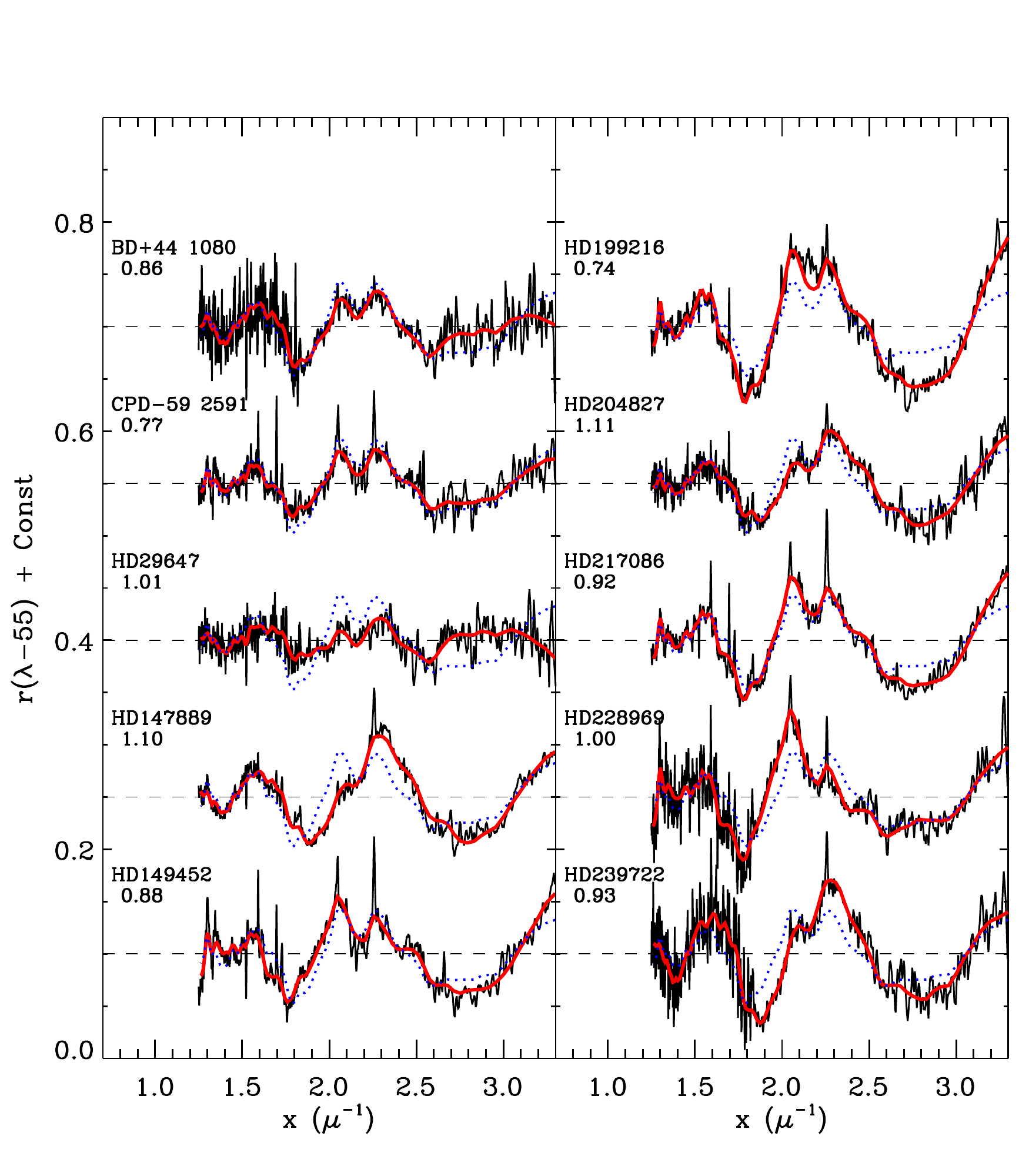}
\end{center}
\vspace{-0.3in}
\caption{Plots of the $r(\lambda-55)$ at 5~\AA\ sampling for the 10 most 
reddened stars in our sample as functions of inverse wavelength.  The name 
of the star and its \ebv\ are given for each curve.  The red curves are  
approximations using the sample mean and the first three PCs.  The sample 
mean (dotted blue) is over plotted on each curve for comparison and the 
dashed lines indicate the zero point for each curve.  The sharp peaks are 
due to the interstellar features listed in Table~\ref{tab:ism}. 
\label{fig:stacks}}
\end{figure*}

\subsection{Wavelength dependence of the ISS variations}
We next examine the systematic variations in the ISS using principal 
component decomposition \cite[see][for a discussion of the use of component 
analysis to characterize extinction]{1980AJ.....85.1651M}.  In this analysis, 
the data were binned even further, to 50~\AA, which is the minimum possible 
sampling for the data, which are smoothed by a 100~\AA\ Gaussian.  The 
result is a set of 99 point spectra.  These were used to construct a  
covariance matrix of excesses.  We use excesses, $E(44-55) r(\lambda-55)$, 
because they give larger weight to the more well defined curves, derived 
from stars with larger reddenings.  The covariance matrix is given by 
\begin{eqnarray}
S_{ij} & = & \frac{1}{N-1}\sum_{n=1}^{N} [r(\lambda_i -55)^{(n)} -\langle 
r(\lambda_i -55)\rangle] E(44-55)^{(n)}  \nonumber \\
& &      \times [r(\lambda_j -55)^{(n)} -\langle r(\lambda_j -55)\rangle] 
         E(44-55)^{(n)}
\end{eqnarray}
where $N = 71$, the number of stars in the sample, $i$ and $j$ vary from 1 
to $N_\lambda = 99$, the number of wavelength points, and 
\begin{equation} 
\langle r(\lambda_i -55)\rangle = \sum_{n=1}^{N} r(\lambda_i -55)^{(n)} 
                                 E(44-55)^{(n)}/\sum_{n=1}^N E(44-55)^{(n)}
\end{equation} 
is the weighted mean.  The $N_\lambda$ eigenvalues, $e_i$, and eigenvectors 
(or principal components, PCs), $v(\lambda)_i$, of this matrix give the 
magnitudes and shapes of the variations, in order of the magnitudes of the 
variances.  We note that the matrix $S_{ij}$ is singular, since there are 
more wavelength points than objects in the sample.  However, this simply 
means that the $N_\lambda -N = 32$ smallest eigenvalues are zero.

A given residual, $r(\lambda-55)^{(n)}$, can be approximated in terms of 
a specified number, $N_{pc}$, of PCs as 
\begin{equation}
r(\lambda-55)^{(n)} \approx \sum_{i=1}^{N_{pc}} \alpha_i^{(n)} v(\lambda)_i 
\end{equation}
where the relation is exact when $N_{pc} = N_\lambda$, and the coefficients 
are given by  
\begin{equation}
\alpha_i^{(n)} = \sum_{k=1}^{N_\lambda} v(\lambda_k)_i r(\lambda_k-55)^{(n)} 
\end{equation}

For any criterion of the significance of the principal components, some 
estimate of the observational errors is required.  For this purpose, we 
adopt $\sigma_{obs}/\sqrt{10} = 0.0118$, where the numerical radical
accounts for the rebinning to 99 points.  This value can be compared to 
the magnitudes of the eigenvalues derived from the sample covariance 
matrix.  The ratio of the square roots of the 5 largest eigenvalues to 
this value are: 4.92, 3.93, 2.26, 1.57 and 1.34.  Consequently, we see 
that the variance along the first 3 components are larger than twice 
the expected mean error, and these are considered significant.  
Furthermore, if we define the fraction of the total variance due to a 
particular component, $e_i$,  as $e_i/\sum_1^{N_\lambda} e_n$, the first 
three components account for 44, 31 and 10\% of the total variance, 
respectively, for a total of 85\%.  In contrast, the fourth component 
accounts for less than 4\%, a value more representative of the expected 
random errors.  Figure~\ref{fig:pcs} shows how the three largest PCs 
affect the mean residual.  The effect of each PC has been scaled by the 
magnitude of its associated eigenvalue.  The first PC appears to describe 
the relative strength of the two peaks near 2.1 \invmic\ relative to the 
minima near 1.8 and 2.8 \invmic.  The major effects of the second PC are 
changing the relative strengths of the two peaks near 2.1 \invmic\ and 
making the minimum near 2.8 \invmic\ weaken in concert with the short 
wavelength 2.1 \invmic\ peak.  The dominant effect of the third component 
is to change the relative strengths of the two minima, weakening the one 
near 1.8 \invmic\ while strengthening the one near 2.8 \invmic.  It is 
difficult to interpret the variability at the shortest wavelengths.  It 
could be due to the wing of another feature with a peak at $\lambda < 
3000$~\AA, or an artifact of the continuum fitting.

Figure~\ref{fig:stacks} shows the residuals of the 10 most reddened stars 
in our sample.  For each star, the plots show: the unsmoothed residuals, the 
representation of the residuals given by the mean and the three largest PCs 
and, the sample mean residual for comparison.  Note that the first three PCs 
produce excellent fits to all of the curves, except for the interstellar 
lines and DIBS listed in Table~\ref{tab:ism}.  This suggests that there may 
be no more than three parameters influencing the ISS.  As expected from 
Figure \ref{fig:kperps}, the features can vary significantly in strength 
(compare HD~29647 and HD~199216), and the strengths of all three peaks vary 
relative to one another (compare HD~147889 and HD~228969).  The feature 
widths also vary, with the feature near 1.6 \invmic\ (which could have 
more than one component) showing the largest variation (compare 
BD$+44^\circ 1080$ and CPD$-59^\circ 2591$).  The position of the broad 
minimum near 1.8 \invmic, which is normally attributed to the VBS, appears 
to shift depending on the relative strengths of the three peaks.  This 
leads us to believe that its position is influenced by the wings of the 
surrounding absorption peaks. 

\section{Modeling the ISS}\label{sec:modeling}
In this section, we describe an {\it ad hoc} model which enables us to 
isolate and quantify the major features seen in the residuals.  Unlike the 
previous section, we use the resampled, but unsmoothed curves. Inspection 
of the mean residual curve shown in Figure~\ref{fig:kperps} and the 
individual residuals shown in Figure~\ref{fig:stacks} suggests that the most 
prominent aspects of the curves can be represented by a smooth background 
extinction and 3 distinct features centered near the strong DIBs at 4428, 
4882, and 6383~\AA.  The upswings in the residuals at each end of the 
wavelength interval are most likely artifacts of the polynomial background 
fit.  We also search for correlations among the parameters used to fit the 
data, and discuss how such correlations might arise.

\subsection{Fitting the curves}  
Each of the three strongest features are modeled as a Drude profile of the 
form given by equation~(\ref{eq:drude}).  Drude functions are appealing 
because they have a solid physical basis and have been shown to provide 
excellent representations of dust extinction \citep{1986ApJ...307..286F} 
and emission \citep[e.g.][]{2001ApJ...551..807D}.  As previously stated, 
a quartic adequately describes the overall shapes of the curves.  
Consequently, we fit each curve with the following function
\begin{equation}
k(\lambda-V) = \sum_{i=1}^3 c_i D(x, x_i, \gamma_i) + \sum_{j=0}^4 d_j x^j
\label{eq:fit}
\end{equation}
Altogether, this model contains 9 free parameters to describe the features: 
three $c_i$, $x_i$, and $\gamma_i$.  These, along with the five $d_j$, were 
determined by non-linear least squares, using the Interactive Data Language 
(IDL) procedure MPFIT developed by \citet{2009ASPC..411..251M}. 
We also use the notation $a(\lambda_i) = c_i/\gamma_i^2$, where $\lambda_i 
= x_i^{-1}$, to denote the amplitude (maxima) of the profiles.  Because all 
of the stars have been fit in the UV as well \citepalias[][]{2019ApJ...886..108F}, 
parameters for the 2175~\AA\ Drude profile are available for each.  
Consequently, we were able to remove the contribution to the curves of the 
long wavelength tail of the 2175~\AA\ Drude profile (although this had 
little influence on the results).  The regions of interstellar lines and 
the strong DIBs listed in Table~\ref{tab:ism} were given zero weight.  In 
contrast, the regions near the Drude peaks were given 9 times the weight of 
the surrounding regions.  This restrains the fitting routine from using the 
tails of the Drude profiles to help the overall agreement of purely 
continuum regions and emphasizes the influence of the positions and widths 
of the profiles \footnote{The regions emphasized by the weights were $1.45 
\leq \mu^{-1} \leq 1.75$ for the position of the long wavelength feature 
and $1.9 \leq \mu^{-1} \leq 2.45$ for both short wavelength positions.  
The values determined by the fitting procedure were well within these 
limits in all cases.}.  We also constrained the $\gamma_i$ to be $\leq 
0.5$ \invmic.  This constraint keeps the Drude profiles from getting so 
wide that they become entangled in the continuum fitting. 

We note that equation~(\ref{eq:fit}) is, in fact, an approximation.  The 
observed curves actually have the form 
\begin{eqnarray}
k(\lambda -55) & = & \frac{E(\lambda-55)_D +E(\lambda-55)_B}{E(44-55)_D +
                    E(44-55)_B}  \label{eq:excess} \\
& \simeq & \left[\frac{E(\lambda-55)_D}{E(44-55)_B} + 
             k(\lambda-55)_B \right] \nonumber \\  
& & \times \left[1 -\frac{E(44-55)_D}{E(44-55)_B} \right] \label{eq:final}
\end{eqnarray}
where the subscripts $D$ and $B$ refer to the Drude and background 
contributions, and equation (\ref{eq:final}) follows since $E(44-55)_D << 
E(44-55)_B$.  The term inside the first set of brackets in 
equation~(\ref{eq:final}) can be rewritten as
\begin{eqnarray}
\frac{E(\lambda-55)_D}{E(44-55)_B} +k(\lambda-55)_B & = & \sum_i \frac{c^0_i 
     }{E(44-55)_B} D(\lambda -55)_i \nonumber \\ 
   & & + k(\lambda-55)_B 
\end{eqnarray}
which has the same form as equation (\ref{eq:fit}).  We see that the 
second term in this equation cannot affect the Drudes and that the 
$c_i$ determined from the fits to equation (\ref{eq:fit}) are actually 
proportional to $c^0_i/E(44-55)_B$.  The second bracketed term in equation 
(\ref{eq:final}) is a multiplicative constant for each line of sight which 
is $\lesssim 1$.  Consequently, it can rescale everything but it cannot 
affect the $x_{0i}$, $\gamma_i$, or the ratios of the $c_i$.  
Consequently, equation (\ref{eq:fit}) is a good approximation, but we need 
to keep in mind that the $c_i$ derived from the fits are actually 
$\simeq c_i^0/E(44-55)_B$, where the $c^0_i$ can be directly related to 
physical parameters described in \S~\ref{sec:relations}. 

Table~\ref{tab:params} lists some properties of the fit parameters for the 
24 stars in our sample with $E(44-55) \geq 0.5$ (values for the entire 
sample are very similar).  For each parameter listed in the first column, 
the Table gives: mean, standard deviation, minimum, maximum, and range, 
defined as $100\times$(max--min)/mean.  It is immediately apparent that the 
$x_i$ values are quite robust and that their total ranges vary by only a 
few percent.  Consequently, we refer to them as the 4370, 4870 and 6800~\AA\ 
features hereafter.  In contrast, all of the other parameters have ranges 
that vary by more than a factor of 2.  

\begin{figure*}
\begin{center}
\includegraphics[width=1.0\linewidth]{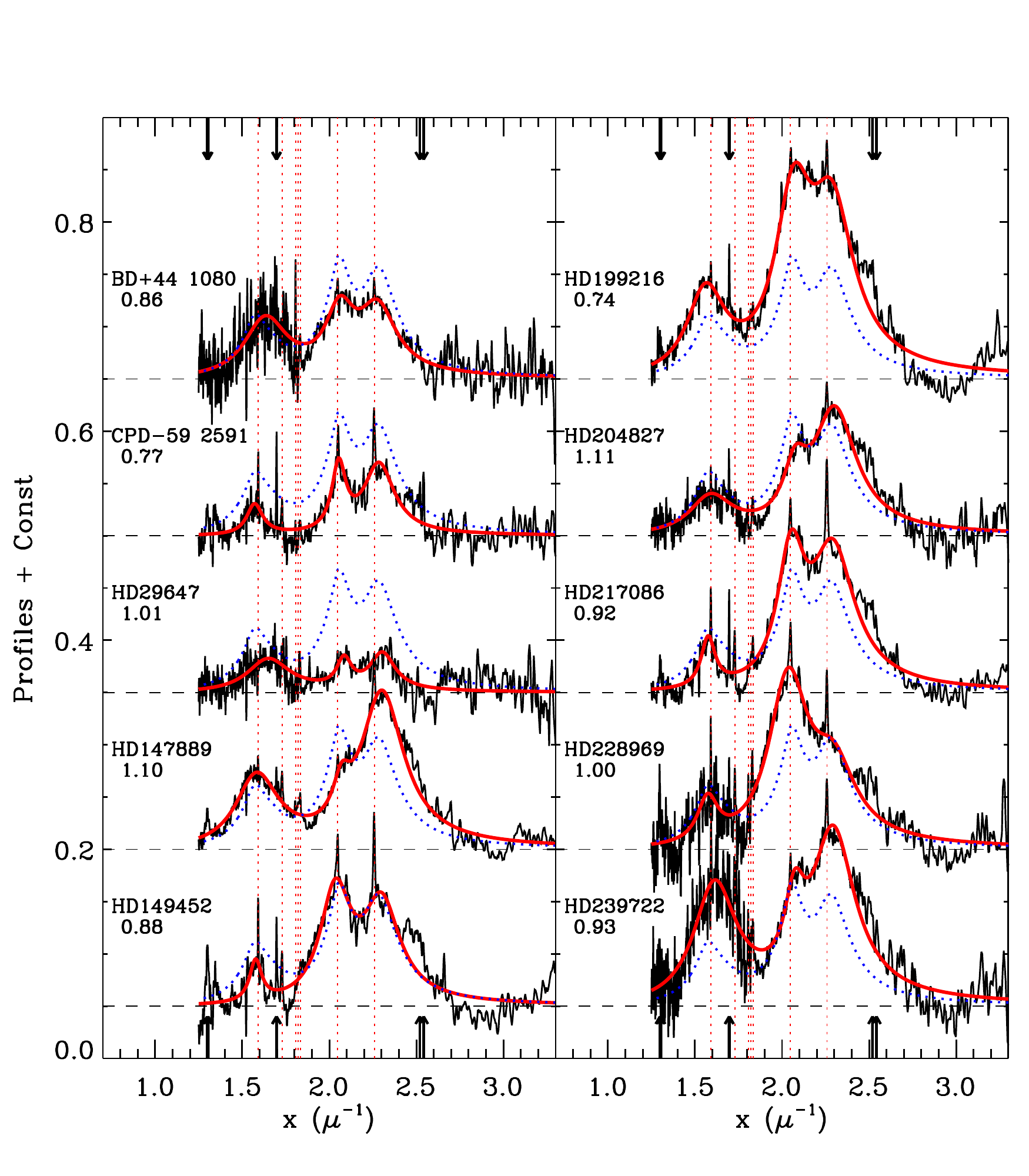} 
\end{center}
\vspace{-.4in}
\caption{Plots of the profile fits for the 10 most reddened stars in our 
sample as functions of inverse wavelength.  The name of the star and its 
\ebv\ are given for each curve.  The thick red curves are the Drude fits, 
the dotted blue curve is the sample mean fit, the black curve is the data 
and the dashed lines give the zero point for each profile.  The arrows 
show the positions of interstellar absorption lines (which appear as 
excess extinction) and the vertical dotted red lines show the positions of 
the strong DIBs listed in Table~\ref{tab:ism}.
\label{fig:fits}}
\end{figure*}

\begin{longrotatetable}
\begin{deluxetable*}{lllllllllllllll}
\tablecaption{Fit Parameters\label{tab:fitparams}}
\tablewidth{700pt}
\tabletypesize{\scriptsize}
\tablehead{\colhead{Star} & \colhead{$a_l \times 10^3$} & \colhead{$a_2 
\times 10^3$} & 
\colhead{$a_3 \times 10^3$} & \colhead{$\gamma_1$} & \colhead{$\gamma_2$} & 
\colhead{$\gamma_3$} & \colhead{$x_{10}$} & \colhead{$x_{20}$} & 
\colhead{$x_{03}$} & \colhead{$b_0$} & \colhead{$b_1$} & \colhead{$b_2$} & 
\colhead{$b_3$} & \colhead{$b_4$}
} 
\startdata
BD+44 1080  &   4.608 &   3.005 &   4.009 &   0.300 &   0.230 &   0.262 &   1.636 &   2.065 &   2.285 &   9.408 & -27.801 &     14.10 &     18.91 &    -17.21 \\ 
BD+56 517   &   2.140 &   6.139 &   6.854 &   0.176 &   0.209 &   0.251 &   1.569 &   2.057 &   2.290 &   5.270 &  -4.389 &    -34.87 &     63.59 &    -32.06 \\ 
BD+56 518   &   8.050 &   7.058 &   8.537 &   0.300 &   0.231 &   0.300 &   1.577 &   2.052 &   2.301 &   6.000 &  -7.353 &    -31.61 &     63.44 &    -32.85 \\ 
BD+56 576   &   1.664 &   6.462 &   3.377 &   0.151 &   0.205 &   0.195 &   1.564 &   2.057 &   2.285 &   5.071 &  -2.996 &    -37.72 &     65.43 &    -32.17 \\ 
BD+69 1231  &   0.423 &   3.985 &   1.087 &   0.083 &   0.182 &   0.109 &   1.573 &   2.068 &   2.281 &  12.913 & -54.559 &     86.42 &    -63.74 &     16.75 \\ 
BD+71 92    &  11.018 &  10.219 &   1.806 &   0.300 &   0.256 &   0.158 &   1.609 &   2.068 &   2.273 &   6.851 &  -8.928 &    -40.25 &     87.38 &    -48.11 \\ 
CPD-41 7715 &   0.228 &   2.048 &   1.225 &   0.086 &   0.171 &   0.148 &   1.571 &   2.052 &   2.290 &   3.684 &   4.846 &    -51.30 &     72.90 &    -32.61 \\ 
CPD-57 3507 &   7.548 &   0.543 &   0.304 &   0.275 &   0.088 &   0.056 &   1.585 &   2.058 &   2.305 &   2.203 &  15.575 &    -78.23 &    100.73 &    -42.57 \\ 
CPD-57 3523 &   0.937 &   1.057 &  10.445 &   0.148 &   0.108 &   0.300 &   1.668 &   2.069 &   2.307 &   4.789 &  -2.619 &    -35.72 &     61.84 &    -30.92 \\ 
CPD-59 2591 &   0.439 &   0.791 &   3.192 &   0.121 &   0.114 &   0.218 &   1.569 &   2.056 &   2.289 &   1.918 &  15.054 &    -72.46 &     92.24 &    -39.88 \\ 
CPD-59 2600 &   0.277 &   1.113 &   4.024 &   0.085 &   0.118 &   0.216 &   1.579 &   2.053 &   2.307 &   3.477 &   4.449 &    -46.34 &     64.44 &    -28.96 \\ 
CPD-59 2625 &   7.622 &   4.809 &   0.015 &   0.300 &   0.232 &   0.010 &   1.621 &   2.040 &   2.236 &   3.714 &   8.139 &    -67.51 &     98.29 &    -45.67 \\ 
HD13338     &   2.480 &   9.761 &   8.707 &   0.174 &   0.230 &   0.271 &   1.561 &   2.059 &   2.284 &   5.317 &  -5.032 &    -33.28 &     62.44 &    -32.01 \\ 
HD14250     &   1.869 &  13.452 &   8.031 &   0.162 &   0.268 &   0.278 &   1.561 &   2.049 &   2.296 &   5.512 &  -5.677 &    -33.09 &     63.22 &    -32.53 \\ 
HD14321     &   1.147 &   4.361 &   4.823 &   0.139 &   0.187 &   0.232 &   1.565 &   2.056 &   2.288 &   6.718 & -14.099 &     -9.58 &     34.12 &    -19.36 \\ 
HD17443     &   0.003 &   0.218 &   2.180 &   0.006 &   0.073 &   0.192 &   1.664 &   2.084 &   2.265 &  15.595 & -70.847 &    124.15 &   -103.36 &     32.81 \\ 
HD18352     &   1.682 &   5.501 &  13.092 &   0.154 &   0.194 &   0.300 &   1.563 &   2.054 &   2.285 &   7.070 & -17.966 &      1.55 &     21.90 &    -14.74 \\ 
HD27778     &   6.604 &   1.408 &   3.910 &   0.300 &   0.152 &   0.224 &   1.622 &   2.072 &   2.269 &   6.868 & -11.659 &    -25.22 &     61.91 &    -34.78 \\ 
HD28475     &   6.932 &   4.309 &   1.314 &   0.300 &   0.178 &   0.127 &   1.570 &   2.061 &   2.286 &   3.113 &   9.944 &    -69.94 &    100.94 &    -46.48 \\ 
HD29647     &   2.711 &   0.249 &   0.656 &   0.300 &   0.100 &   0.142 &   1.628 &   2.093 &   2.314 &   8.650 & -26.976 &     24.81 &     -7.42 &     -1.38 \\ 
HD30122     &  10.780 &   3.551 &  12.304 &   0.300 &   0.169 &   0.300 &   1.578 &   2.072 &   2.289 &   2.382 &  14.026 &    -80.90 &    115.66 &    -53.95 \\ 
HD30675     &   9.493 &   0.307 &  11.180 &   0.300 &   0.092 &   0.288 &   1.625 &   2.079 &   2.302 &   3.324 &   8.898 &    -69.30 &    102.72 &    -48.55 \\ 
HD37061     &   0.080 &   0.060 &   0.048 &   0.056 &   0.035 &   0.051 &   1.545 &   2.050 &   2.306 &   2.082 &  14.118 &    -66.36 &     78.32 &    -30.74 \\ 
HD38087     &   0.468 &  13.017 &   0.331 &   0.101 &   0.300 &   0.085 &   1.565 &   2.005 &   2.281 &   1.108 &  24.079 &   -103.18 &    133.32 &    -59.14 \\ 
HD40893     &   6.185 &   3.415 &  14.550 &   0.255 &   0.160 &   0.300 &   1.593 &   2.053 &   2.294 &   4.220 &   2.928 &    -56.07 &     91.87 &    -45.83 \\ 
HD46106     &   1.137 &   5.244 &   5.596 &   0.132 &   0.200 &   0.229 &   1.566 &   2.051 &   2.291 &   5.836 &  -9.891 &    -15.82 &     36.12 &    -18.25 \\ 
HD46660     &   3.057 &   2.170 &   8.377 &   0.237 &   0.146 &   0.275 &   1.585 &   2.054 &   2.304 &   5.574 &  -7.491 &    -24.70 &     50.10 &    -25.77 \\ 
HD54439     &   5.813 &   1.690 &  10.967 &   0.288 &   0.119 &   0.300 &   1.594 &   2.062 &   2.308 &   6.653 & -14.204 &     -8.62 &     32.39 &    -17.83 \\ 
HD62542     &   5.204 &   0.590 &   2.163 &   0.300 &   0.116 &   0.253 &   1.679 &   2.057 &   2.280 &   7.013 & -10.168 &    -32.61 &     72.73 &    -40.38 \\ 
HD68633     &   0.569 &   4.630 &   0.607 &   0.118 &   0.280 &   0.115 &   1.555 &   2.014 &   2.301 &   2.569 &  13.145 &    -72.42 &     95.08 &    -41.13 \\ 
HD70614     &   1.230 &   8.159 &   1.201 &   0.137 &   0.237 &   0.150 &   1.562 &   2.034 &   2.287 &   1.234 &  22.737 &   -100.07 &    130.06 &    -56.67 \\ 
HD91983     &   2.926 &   0.521 &  10.432 &   0.247 &   0.076 &   0.300 &   1.642 &   2.080 &   2.281 &   8.464 & -29.332 &     34.71 &    -18.44 &      2.71 \\ 
HD92044     &   3.435 &   2.221 &  11.762 &   0.251 &   0.163 &   0.300 &   1.597 &   2.069 &   2.292 &   6.795 & -19.126 &     13.28 &     -1.02 &     -1.94 \\ 
HD93028     &   0.344 &   1.130 &   1.966 &   0.087 &   0.108 &   0.146 &   1.562 &   2.055 &   2.283 &   5.167 &  -5.312 &    -27.06 &     50.09 &    -26.23 \\ 
HD93222     &   0.002 &   0.842 &   4.427 &   0.004 &   0.108 &   0.222 &   1.593 &   2.061 &   2.308 &   5.783 & -15.826 &     14.78 &    -11.54 &      4.41 \\ 
HD104705    &   0.113 &   0.776 &   8.002 &   0.049 &   0.104 &   0.281 &   1.594 &   2.062 &   2.289 &   6.839 & -18.225 &      7.30 &     11.19 &     -9.15 \\ 
HD110336    &   4.604 &   0.476 &   2.660 &   0.264 &   0.111 &   0.182 &   1.580 &   2.092 &   2.294 &   6.591 & -15.057 &     -2.17 &     20.12 &    -11.60 \\ 
HD110946    &   8.130 &   4.056 &   6.718 &   0.300 &   0.188 &   0.300 &   1.567 &   2.057 &   2.267 &   4.475 &   1.333 &    -48.36 &     76.06 &    -35.86 \\ 
HD112607    &   0.255 &   1.595 &   1.861 &   0.081 &   0.158 &   0.146 &   1.566 &   2.072 &   2.278 &   7.202 & -18.444 &      3.87 &     17.13 &    -12.11 \\ 
HD142096    &   0.488 &   0.189 &   0.015 &   0.085 &   0.045 &   0.010 &   1.550 &   2.055 &   2.295 &   4.061 &   3.028 &    -47.11 &     67.80 &    -30.44 \\ 
HD142165    &   0.034 &   0.386 &   0.243 &   0.023 &   0.060 &   0.054 &   1.550 &   2.055 &   2.288 &  15.724 & -78.050 &    159.93 &   -162.83 &     64.44 \\ 
HD146285    &   0.146 &   7.655 &   0.214 &   0.070 &   0.283 &   0.067 &   1.561 &   2.033 &   2.285 &   7.644 & -20.264 &      7.74 &     11.79 &     -9.42 \\ 
HD147196    &   1.997 &   0.004 &   4.262 &   0.158 &   0.007 &   0.300 &   1.655 &   2.073 &   2.171 &  10.089 & -39.121 &     61.55 &    -56.01 &     22.40 \\ 
HD147889    &   6.211 &   0.592 &  13.005 &   0.300 &   0.129 &   0.300 &   1.583 &   2.076 &   2.305 &   1.700 &  13.447 &    -63.59 &     78.67 &    -33.07 \\ 
HD149452    &   0.332 &   6.188 &   7.209 &   0.093 &   0.239 &   0.277 &   1.575 &   2.034 &   2.308 &   3.698 &   5.096 &    -54.50 &     80.27 &    -37.41 \\ 
HD164073    &   1.019 &   6.958 &   0.012 &   0.114 &   0.295 &   0.011 &   1.592 &   1.950 &   2.325 &   4.140 &   4.783 &    -55.60 &     80.46 &    -37.24 \\ 
HD172140    &   3.504 &   0.706 &   8.434 &   0.226 &   0.087 &   0.231 &   1.577 &   2.058 &   2.303 &   4.687 &  -6.024 &    -16.05 &     26.81 &    -10.86 \\ 
HD193322    &   0.378 &   6.065 &   1.764 &   0.099 &   0.230 &   0.154 &   1.563 &   2.038 &   2.294 &   5.682 &  -3.931 &    -42.85 &     78.98 &    -41.00 \\ 
HD197512    &   3.506 &   7.942 &   6.571 &   0.207 &   0.224 &   0.253 &   1.562 &   2.064 &   2.276 &   5.575 &  -6.276 &    -31.69 &     62.14 &    -32.22 \\ 
HD197702    &   0.910 &   7.500 &  16.113 &   0.112 &   0.222 &   0.300 &   1.601 &   2.074 &   2.285 &   9.736 & -36.463 &     47.57 &    -26.77 &      3.71 \\ 
HD198781    &   0.192 &   1.661 &   8.966 &   0.077 &   0.140 &   0.300 &   1.553 &   2.050 &   2.274 &   6.670 & -17.209 &      5.45 &     10.85 &     -7.69 \\ 
HD199216    &   6.576 &  11.963 &  12.661 &   0.278 &   0.272 &   0.300 &   1.562 &   2.069 &   2.295 &   7.772 & -22.748 &     12.66 &     10.66 &    -10.44 \\ 
HD204827    &   3.547 &   1.745 &  10.317 &   0.300 &   0.183 &   0.300 &   1.591 &   2.079 &   2.307 &  10.212 & -37.766 &     48.84 &    -28.49 &      5.28 \\ 
HD210072    &   7.334 &   2.680 &   4.819 &   0.300 &   0.169 &   0.251 &   1.616 &   2.048 &   2.269 &  10.181 & -33.035 &     27.94 &      1.98 &     -9.11 \\ 
HD210121    &   0.088 &   0.452 &   1.629 &   0.059 &   0.108 &   0.164 &   1.546 &   2.079 &   2.286 &  14.630 & -64.916 &    112.03 &    -94.18 &     30.96 \\ 
HD217086    &   0.605 &   5.127 &  11.475 &   0.113 &   0.204 &   0.300 &   1.573 &   2.052 &   2.297 &   5.083 &  -4.333 &    -32.83 &     59.93 &    -30.56 \\ 
HD220057    &   5.044 &   1.226 &   1.557 &   0.241 &   0.121 &   0.141 &   1.566 &   2.064 &   2.293 &   6.119 &  -9.277 &    -21.49 &     45.39 &    -22.93 \\ 
HD228969    &   1.390 &  15.484 &   5.905 &   0.162 &   0.300 &   0.300 &   1.577 &   2.033 &   2.304 &   4.088 &   8.469 &    -78.76 &    123.01 &    -59.90 \\ 
HD236960    &  10.749 &   5.502 &  13.059 &   0.300 &   0.206 &   0.300 &   1.591 &   2.058 &   2.293 &   4.703 &  -1.371 &    -41.20 &     69.07 &    -33.43 \\ 
HD239693    &  10.237 &   3.043 &  10.555 &   0.300 &   0.174 &   0.299 &   1.569 &   2.064 &   2.294 &   3.735 &   5.604 &    -59.06 &     88.99 &    -41.62 \\ 
HD239722    &   9.571 &   1.526 &  14.170 &   0.300 &   0.147 &   0.300 &   1.619 &   2.077 &   2.298 &   5.025 &  -1.484 &    -46.51 &     81.58 &    -41.37 \\ 
HD239745    &   1.359 &   7.758 &  10.689 &   0.153 &   0.245 &   0.300 &   1.561 &   2.057 &   2.294 &   5.649 &  -8.412 &    -22.11 &     47.34 &    -24.92 \\ 
HD282485    &   3.507 &   2.253 &   9.103 &   0.222 &   0.166 &   0.249 &   1.616 &   2.085 &   2.303 &   6.738 & -15.299 &     -7.34 &     34.74 &    -21.25 \\ 
HD292167    &   0.387 &   2.690 &  14.033 &   0.081 &   0.158 &   0.300 &   1.588 &   2.053 &   2.297 &   5.649 & -10.217 &    -13.57 &     34.76 &    -19.09 \\ 
HD294264    &   3.908 &   7.970 &   0.006 &   0.300 &   0.300 &   0.012 &   1.605 &   1.989 &   2.290 &  -2.057 &  43.472 &   -144.91 &    171.47 &    -72.25 \\ 
HD303068    &  10.141 &   7.196 &   5.626 &   0.300 &   0.260 &   0.248 &   1.651 &   2.061 &   2.323 &   2.078 &  21.228 &   -107.95 &    152.48 &    -71.17 \\ 
NGC 2244 11 &   1.543 &   2.904 &  10.803 &   0.148 &   0.151 &   0.280 &   1.565 &   2.056 &   2.298 &   5.715 & -10.921 &     -9.84 &     27.09 &    -13.89 \\ 
NGC 2244 23 &   8.349 &   2.255 &   5.931 &   0.300 &   0.156 &   0.247 &   1.594 &   2.058 &   2.292 &   2.755 &  13.427 &    -80.73 &    114.42 &    -52.42 \\ 
Trumpler 14 &   0.666 &   0.530 &   7.158 &   0.122 &   0.097 &   0.300 &   1.604 &   2.065 &   2.289 &   3.506 &   3.749 &    -43.21 &     59.35 &    -25.85 \\ 
Trumpler 14 &   1.215 &   1.883 &   5.946 &   0.181 &   0.167 &   0.300 &   1.599 &   2.068 &   2.309 &   4.253 &   0.117 &    -37.12 &     55.30 &    -25.09 \\ 
VSS VIII-10 &   7.541 &   6.278 &   3.586 &   0.300 &   0.300 &   0.300 &   1.678 &   1.960 &   2.194 &   6.582 &  -6.723 &    -40.22 &     77.67 &    -40.98 \\ 
\enddata
\end{deluxetable*}
\end{longrotatetable}

Figure~\ref{fig:fits} shows the profile portion of the fits, i.e., the fit 
minus the quartic contribution, for the 10 most reddened stars in our 
sample. The overall quality of the fits is good, but it is also apparent 
that some aspects of the curves are not fit particularly well by our simple 
model (e.g., the blue edge of the curves and the blue wing of the 4370~\AA\ 
profile).  While all the features are, in general, rather weak, their 
strengths vary significantly, from $\lesssim 0.05$ in HD~29647, to $\gtrsim 
0.20$ in HD~199216 and HD~239722, where the first has a very weak 2175~\AA\ 
bump and the latter two quite strong ones \citepalias[see][]{2019ApJ...886..108F}.   

\begin{figure}
\begin{center}
\includegraphics[width=0.9\linewidth]{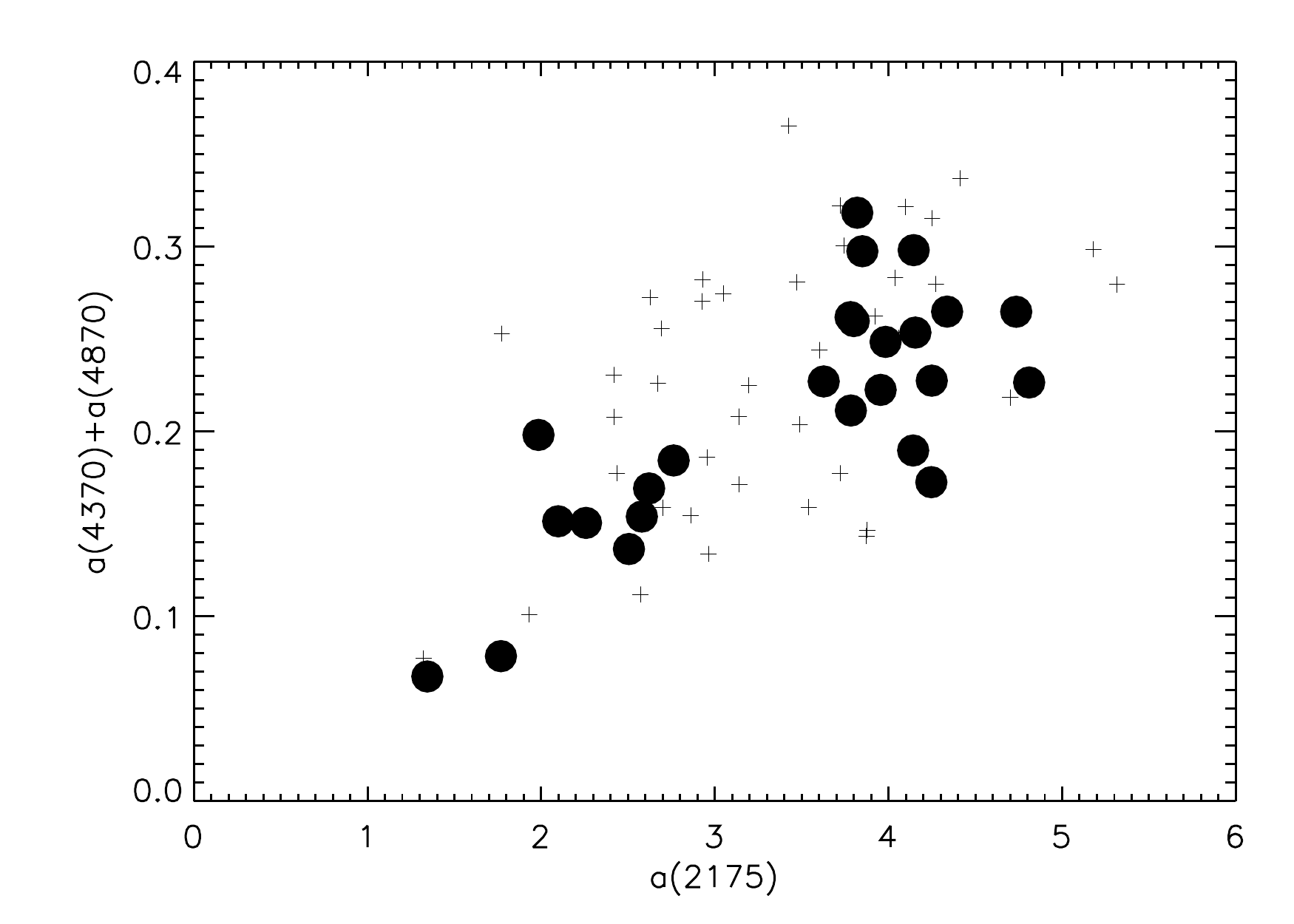} 
\end{center}
\vspace{-0.1in}
\caption{Combined amplitudes of the 4370 and 4870~\AA\ features plotted 
against the amplitude of the 2175~\AA\ UV bump.  Large, filled symbols are 
for values derived from curves with $E(44-55) \geq 0.5$ mag, which should 
be most accurate.  
\label{fig:fitbump}}
\end{figure}

Plots of correlations among the individual fit parameters show that the 
amplitudes of the features, $a(4370)$ and $a(4870)$, correlate well with 
$a(2175)$, but that $a(6300)$ does not.  Figure~\ref{fig:fitbump} shows the 
correlation between $a(4370) + a(4870)$ and $a(2175)$.  Although there is a 
clear correlation between these two quantities, the relation is not quite 
as strong as the one between the RMS of the residuals and $a(2175)$ shown 
in Figure~\ref{fig:bump}.  This is, no doubt, because our simple model does 
not capture all of the more complex structure that is related to $a(2175)$ 
and because the derived parameters may be influenced by the continuum fit.  

\subsection{Relations among the fit parameters\label{sec:relations}}  
\begin{figure}
\begin{center}
\includegraphics[width=1.\linewidth]{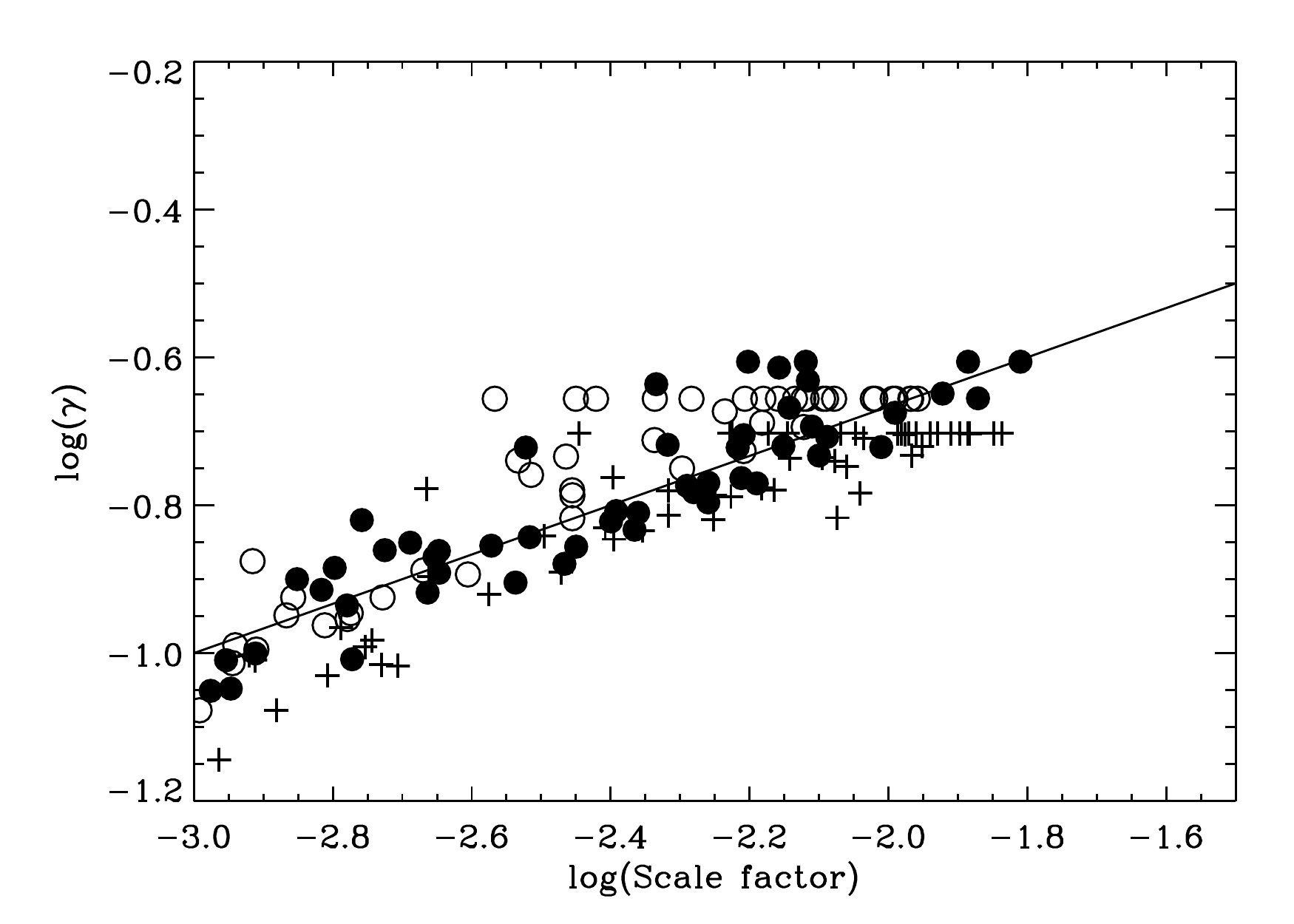}
\end{center}
\vspace{-0.2in}
\caption{Log-log plot of the $\gamma$s versus the scale factors for all of 
the features.  Symbols are: crosses for 4370~\AA, filled circles for 4870~\AA\ 
and open circles for 6300~\AA. The line has a slope of 3, which represents a 
cubic.  To align the data, a single constant was added to all of the $\log 
\gamma$ data for each wavelength.  Note that a few of the $\gamma$s for the 
6300 and 4870~\AA\ fits ran into the upper limit in the non-linear least 
squares routine, causing a horizontal row of points near $\log \gamma = 
-0.65$. 
\label{fig:llall}}
\end{figure}

Since 3 PCs can explain 85\% of the variability, there should be several 
constraints among the 9 fit parameters used to fit the three profiles.  
Three constraints can be found by fixing the central positions of the three 
$x_{0i}$, since their variance is so small.  When this is done, the quality 
of the fits is hardly affected.  We searched for additional relations 
among the other parameters and noticed that the scale factors, $c_i$ and the 
widths, $\gamma_i$ are related.  Figure~\ref{fig:llall} shows the $c_i - 
\gamma_i$ relations for all three features.  A different constant for each 
feature has been added to the $\log \gamma$, vertically shifting the points, 
so that each feature can be seen to follow a simple linear relation, which 
has a slope 3.  Notice that several of the points for the 6630~\AA\ feature 
and a few for the 4370~\AA\ one, form horizontal lines near $\log \gamma 
\simeq -0.6$.  This is due to the constraint enforced on $\gamma_i$ 
in the fitting routine described above.  Nevertheless, over most of the 
range, the vast majority of the points fall along a line, which implies that 
there are constants, $A_i$, such that 
\begin{equation}
c_i = \frac{c_i^0}{E(44-55)_B} = A_i \gamma_i^3 \;\;, 
\label{eq:relate}
\end{equation} 
where $A_i = 0.35, 0.70,$ and 0.45, for the 6300, 4870, and 4370~\AA\ 
features, respectively.  These relations can be used to eliminate the $c_i$, 
leaving the $\gamma_i$ as the only free parameters in the fits.  When this 
was done, the fits degraded significantly.  However, when the constraints 
were only applied to the 4870 and 4370~\AA\ features, and both $c$ and 
$\gamma$ were allowed to be free for 6300~\AA, the quality of the fits 
became comparable to that obtained when no constraints are applied.  As a 
result, we can find 5 constraints, implying that only 4 free parameters are 
needed to fit all three optical features.  

\begin{figure}
\begin{center}
\includegraphics[width=1.\linewidth]{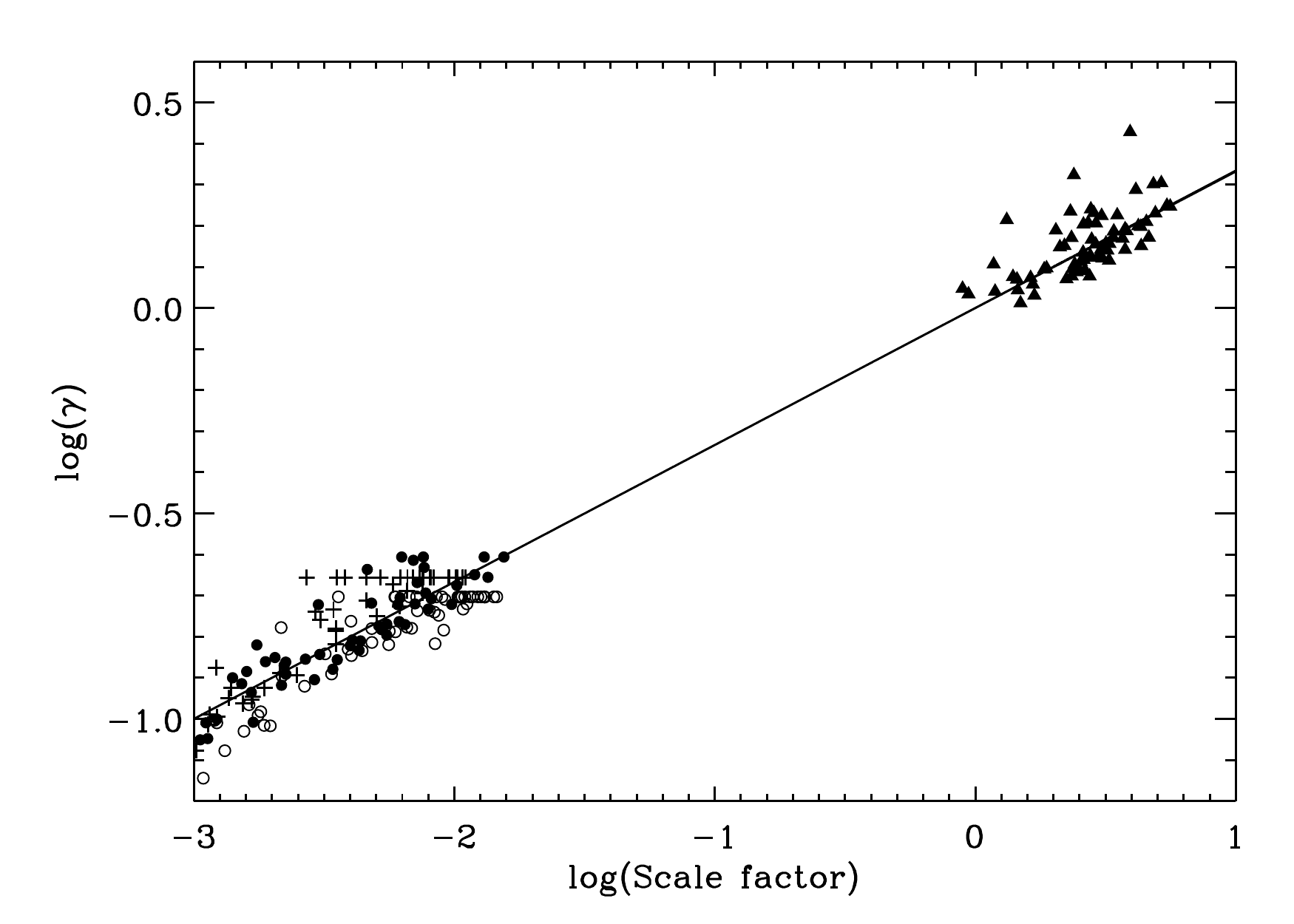}
\end{center}
\vspace{-.2in}
\caption{Log-log plot of the $\gamma$s versus the scale factors for all of 
the features {\it including} the UV 2175~\AA\ bump.  Symbols for the optical 
features are the same as in Figure \ref{fig:llall} and the triangles 
represent the 2175~\AA\ data.  As in Figure \ref{fig:llall}, the line has a 
slope of 3 and the a single constant was added to all of the $\log \gamma$ 
data for each feature.  
\label{fig:plus2175}}
\end{figure}

The question that naturally arises is whether the parameters of the 2175~\AA\ 
bump follow the same relation.  Figure~\ref{fig:plus2175} extends the 
range shown in Figure~\ref{fig:llall}, to include the 2175~\AA\ bump.  In 
this case, $A = 4.0$.  Although there is considerable scatter, it appears 
that the lower envelope of the 2175~\AA\ parameters also follows a line of 
slope 3.  

We now consider what the $c \propto \gamma^3$ relations might imply.  To 
begin, we adopt a very simplistic model which consists of one grain size 
of Drude absorbers and one grain size of background absorbers along each 
line of sight.  For a single population of Drude particles, $N_D \propto 
c^0/\gamma$, where $N_D$ is the number of absorbers \citep[see][]{1983asls.book.....B}.
We also assume that $N_B \propto E(B-V)_B$, where $N_B$ is the number of 
background absorbers.  As a result, equation(\ref{eq:relate}) becomes 
$N_D/N_B \propto \gamma^2$.  In the Drude model, $\gamma \propto 1/r_D$, 
where $r_D$ is the radius of the absorbers and their geometric cross 
section is $a_D \propto r_D^2$.  This gives $N_D/N_B \propto a_D^{-1}$. 
If we also assume that $N_B$ is fixed, then the previous relation implies 
that when the Drude grains are small, there are relatively more of them 
along the line of sight compared to background absorbers and when they are 
large, there are relatively fewer.  In addition, the total mass in Drude 
grains, $M_D$, is related to $r_D$ by $M_D \propto \rho N_D r_{D}^3$.  
Utilizing the previous result, that $N_{D} \propto r_D^{-2}$\ implies that 
$M_D \propto r_D$, i.e., there is less mass in the Drude grains when they are 
smaller.  Thus, the empirical relations given by eq.\ (\ref{eq:relate}) 
imply that the relative number of Drude absorbing grains may be similar 
along all Milky Way lines of sight, but that less total mass resides in the 
Drude grains when they are smaller.  We emphasize that this model is 
extremely simplistic and only meant to illustrate how the constraints could 
influence a more comprehensive grain model.  

\begin{figure}
\begin{center}
\includegraphics[width=0.9\linewidth]{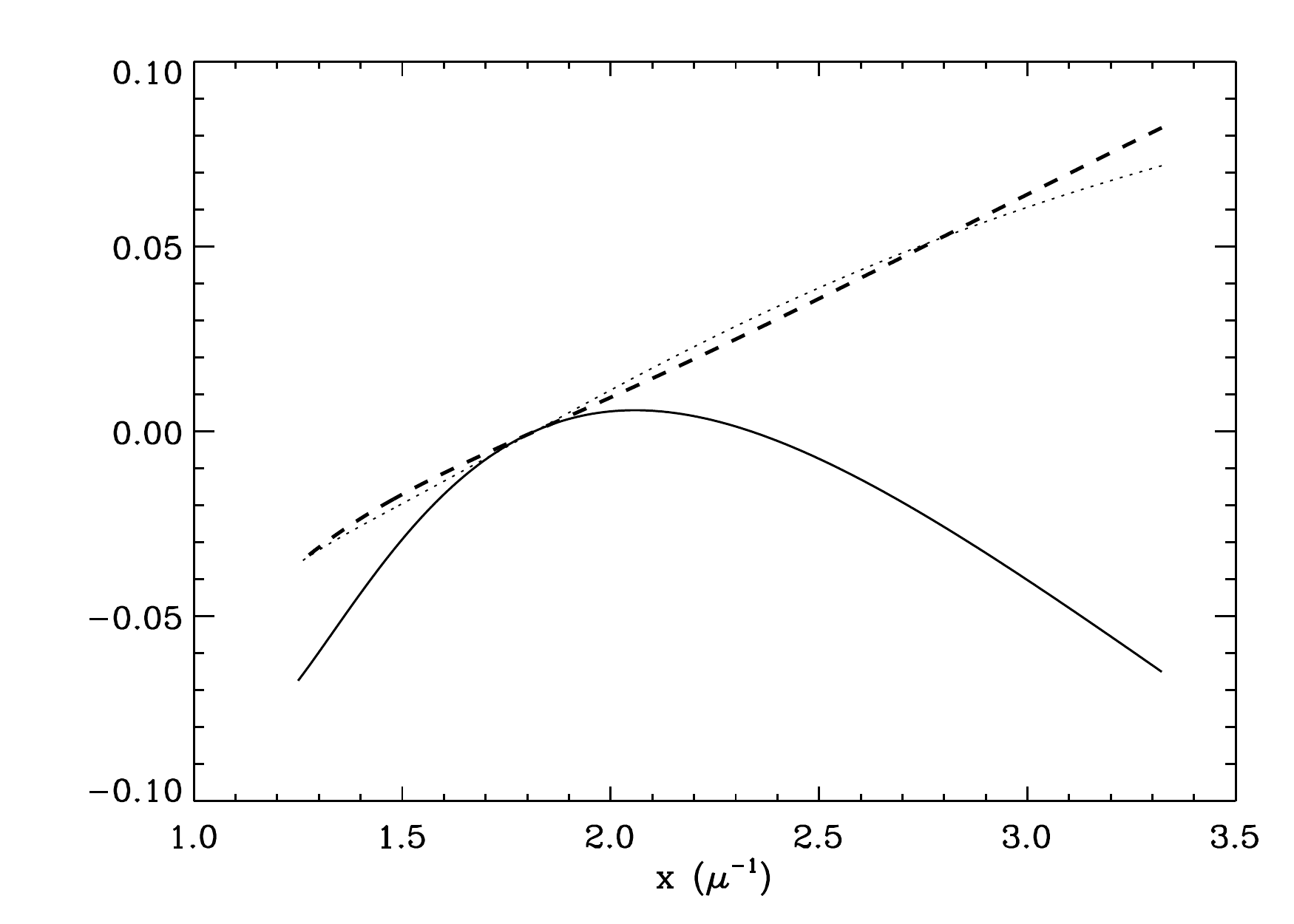}\vfill
\includegraphics[width=0.9\linewidth]{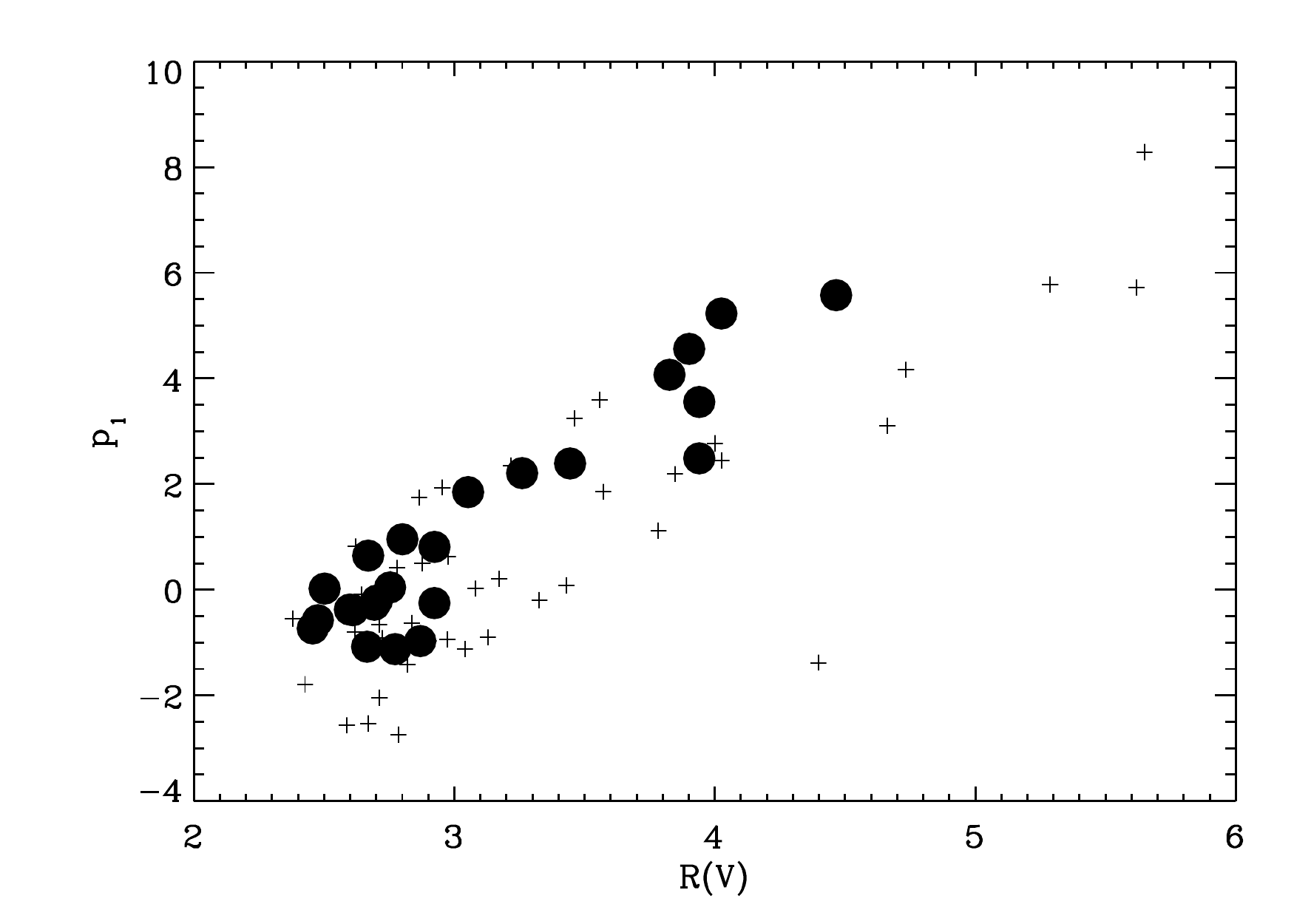}
\end{center}
\vspace{-0.2in}
\caption{Top: The largest (solid) and second largest (dashed) principal 
components of the continuum fits, compared to a scaled version of the 
mean curve (dotted). Bottom: The components of the largest PC plotted 
against \rv.  Large, filled symbols are for values derived from curves 
with $E(44-55) \geq 0.5$ mag.
\label{fig:cont}}
\end{figure}

We also performed principal component analysis on the continuum polynomials 
derived for each curve.  The two largest PCs account for nearly 90\% of the 
continuum variation, 51 and 37\%, respectively.  The upper plot in 
Figure~\ref{fig:cont} shows the wavelength dependence of the two largest 
PCs, as well as a scaled version of the mean polynomial fit.  The lower plot 
shows the relationship between the coefficients of the largest PC and \rv.  
The excellent correlation should not be surprising, since it is well known 
that curves which ``roll over'' in the optical, i.e., are strongly 
influenced by the shape of the largest PC, are typically associated with 
large $R(V)$ values.  We note that the maximum of the largest PC lies very 
near the maximum in the $d k(\lambda-55)/d R(55)$ curve given by 
\citetalias{2019ApJ...886..108F}.  However, it is shortward of the maximum 
in the second PC derived by \cite{2016ApJ...821...78S}.  This difference is 
most likely the result of the very different wavelength baselines used.  

\vfill
\begin{table}
\begin{center}
\caption{Parameter properties}
\begin{tabular}{cccllr}
\label{tab:params}
Name        &  Mean & $\sigma$ & Min   & Max   & Range\\ \hline
            & (mag) & (mag)    & (mag) & (mag) & (\%) \\ 
$a(4370)$   & 0.102 & 0.043    & 0.018 & 0.188 & 166  \\
$a(4870)$   & 0.107 & 0.046    & 0.030 & 0.194 & 153 \\
$a(6300)$   & 0.077 & 0.037    & 0.023 & 0.170 & 190 \\ \hline
            & ( \invmic) & ( \invmic)    & ( \invmic) & ( \invmic) & (\%) \\ 
$x_1$       & 2.288 & 0.012    & 2.26  & 2.31  & 2.2 \\
$x_2$       & 2.054 & 0.013    & 2.08  & 2.01  & 3.6 \\
$x_3$       & 1.587 & 0.027    & 1.66  & 1.55  & 6.7 \\ 
$\gamma_1$  & 0.243 & 0.011    & 0.019 & 0.400 & 156 \\
$\gamma_2$  & 0.179 & 0.007    & 0.030 & 0.288 & 144 \\
$\gamma_3$  & 0.243 & 0.148    & 0.080 & 0.400 & 132 \\ \hline
\end{tabular}
\end{center}
\end{table}

\section{Summary}\label{sec:summary}

We have verified the reality of ISS features in extinction curves by 
demonstrating two properties.  First, its strength is independent of the 
physical parameters of the star used to create a curve.  This can be seen 
by comparing the features for HD~147889 (B2 V), and HD~149452 (O8 V) in 
Figures~\ref{fig:stacks} and \ref{fig:fits}.  Second, the strength of the 
ISS correlates with an interstellar feature, the strength of the 2175~\AA\ 
bump.  Further, as with the 2175 \AA\ feature, curves of stars in the same 
region tend to have the same ISS strength.  We then proceeded to determine 
the magnitude, wavelength dependence and variability of the ISS in two 
different ways.  

In the first approach, we examined the residuals of the extinction curves 
relative to a smooth background that is modeled by a quartic.  It was shown 
that the residual curves have three strong peaks whose relative strengths 
and widths vary substantially from one sight line to another, but whose 
central positions are relatively stable and located near the strong DIBs at 
4428, 4882 and 6284~\AA, suggesting a possible relation.  It was also 
apparent that the VBS is actually a local minimum between absorption 
features.  Next, we demonstrated that the magnitudes of the root mean square 
of the residuals -- which give a crude measure of the strength of the ISS 
-- are poorly correlated with \rv, but strongly correlated with the strength 
of the 2175~\AA\ bump.  Finally, principal component analysis was used to 
reveal that most of the ISS variations can be captured by the first three 
PCs of the residuals, suggesting that as few as three physical parameters 
might be able to explain the variations.

In the second approach, we modeled the ISS with three Drude profiles in 
order to quantify its major structures.  This simple model provides a 
reasonable fit to the ISS, although it has some small inadequacies.  The 
fits showed that the feature locations are very stable, near $\lambda = 
4370$, 4870 and 6300~\AA\, and that the features contribute anywhere from a 
few percent to 20\% of the extinction in those regions.  Further analysis 

showed that the strengths of the 4370 and 4870~\AA\ features correlate with 
the strength of the 2175~\AA\ UV bump, but the 6300~\AA\ feature does not, 
and that none of the features correlate with \rv.  We found relations among 
the model parameters which can reduce them from 9 to 4, without a major 
loss of accuracy.  Interpretation of these relations in terms of a simple 
dust model suggests that the density of dust producing the ISS is similar 
along all lines of sight, but distributed in different grain sizes.  In 
addition, we verified that the strongest variation in the continuum 
extinction is the curvature, and that its strength is related to \rv. 

While it is hoped that our results will be useful for developing physical 
models of dust grains, they also have practical importance.  Although the 
ISS is small, it should be taken into consideration if one wishes to achieve 
accurate fits to the SEDs of reddened stars.  Also, the relation between 
the ISS features and the strength of the 2175~\AA\ bump could be useful for 
estimating the strength of the 2175~\AA\ bump in objects whose optical SEDs 
are either intrinsically smooth or well modeled but whose UV SEDs are poorly 
determined.  

\acknowledgments
Support for program \#13760 was provided by NASA through a grant from 
the Space Telescope Science Institute, which is operated by the Association 
of Universities for Research in Astronomy, Inc., under NASA contract NAS 
5-26555.  We also thank the referee for a thorough and insightful review of 
the manuscript, which improved the final version of the paper.  

\bibliographystyle{aasjournal}

\end{document}